\documentclass[sigconf]{acmart}

\usepackage{multirow}
\usepackage{algorithm}
\usepackage{algorithmic}

\usepackage{microtype}
\usepackage{amsmath}
\usepackage{amsthm}
\usepackage{graphicx}
\usepackage{booktabs}
\usepackage{multirow}

\usepackage{array}
\usepackage{arydshln}
\usepackage{color}
\usepackage{balance}

\newenvironment{enumeratesquish}[2]{\begin{list}{\labelenumi}{\setlength{\itemsep}{#1}\setlength{\labelwidth}{#2}\setlength{\leftmargin}{\labelwidth}\addtolength{\leftmargin}{\labelsep}}}{\end{list}}

\AtBeginDocument{%
  }

\copyrightyear{2025} 
\acmYear{2025} 
\setcopyright{cc}
\setcctype{by}
\acmConference[KDD '25]{Proceedings of the 31st ACM SIGKDD Conference on Knowledge Discovery and Data Mining V.2}{August 3--7, 2025}{Toronto, ON, Canada}
\acmBooktitle{Proceedings of the 31st ACM SIGKDD Conference on Knowledge Discovery and Data Mining V.2 (KDD '25), August 3--7, 2025, Toronto, ON, Canada}
\acmDOI{10.1145/3711896.3737005}
\acmISBN{979-8-4007-1454-2/2025/08}

\begin{document}

\title{Pre-training Time Series Models with Stock Data Customization}

\author{Mengyu Wang}
 \orcid{0009-0002-4160-3791}
 \affiliation{%
 \institution{University of Edinburgh}
 \department{School of Informatics}
 \city{Edinburgh}
 \country{United Kingdom}}
 \email{mengyu.wang@ed.ac.uk}

 \author{Tiejun Ma}
 \orcid{0000-0001-5545-6978}
 \affiliation{%
 \institution{University of Edinburgh}
 \department{School of Informatics}
 \city{Edinburgh}
 \country{United Kingdom}}
 \email{tiejun.ma@ed.ac.uk}

 \author{Shay B. Cohen}
 \orcid{0000-0003-4753-8353}
 \affiliation{%
 \institution{University of Edinburgh}
 \department{School of Informatics}
 \city{Edinburgh}
 \country{United Kingdom}}
 \email{scohen@inf.ed.ac.uk}

\renewcommand{\shortauthors}{Mengyu Wang, Tiejun Ma, \& Shay B. Cohen}

\begin{abstract}
Stock selection, which aims to predict stock prices and identify the most profitable ones, is a crucial task in finance. While existing methods primarily focus on developing model structures and building graphs for improved selection, pre-training strategies remain underexplored in this domain. Current stock series pre-training follows methods from other areas without adapting to the unique characteristics of financial data, particularly overlooking stock-specific contextual information and the non-stationary nature of stock prices. Consequently, the latent statistical features inherent in stock data are underutilized. In this paper, we propose three novel pre-training tasks tailored to stock data characteristics: stock code classification, stock sector classification, and moving average prediction. We develop the Stock Specialized Pre-trained Transformer (SSPT) based on a two-layer transformer architecture. Extensive experimental results validate the effectiveness of our pre-training methods and provide detailed guidance on their application. Evaluations on five stock datasets, including four markets and two time periods, demonstrate that SSPT consistently outperforms the market and existing methods in terms of both cumulative investment return ratio and Sharpe ratio. Additionally, our experiments on simulated data investigate the underlying mechanisms of our methods, providing insights into understanding price series. Our code is publicly available at: https://github.com/astudentuser/Pre-training-Time-Series-Models-with-Stock-Data-Customization.
\end{abstract}



\maketitle

\section{Introduction}
Stock selection is a critical aspect of investment decision-making in the vast stock market~\cite{fan2001stock, sawhney2021stock}. Predicting stock trends to identify the most profitable investment opportunities has become a popular research topic~\cite{preethi2012stock, feng2019temporal, zhang2024reinforcement}. Although financial time series are volatile, they are not entirely random, persistent inefficiencies in markets lead to exploitable patterns~\cite{jegadeesh1993returns, lo1990contrarian, wang2024modeling}. Stock movements are driven by a range of economic and behavioral factors, including volatility dynamics, momentum effects, and sector-specific trends~\cite{eugene1992cross, moskowitz1999industries}. These factors introduce subtle yet recurring signals in the data, which cannot be adequately captured through supervised learning alone. Many advanced methods in financial area are fundamentally focused on uncovering such latent structures to enhance predictive performance~\cite{yoo2021accurate, feng2019temporal}. Thus, the belief that financial markets contain learnable, non-random signals is a foundational assumption shared across both academic and applied finance communities. 

Traditional approaches often rely on time-series analysis models to evaluate stock price data~\cite{adebiyi2014comparison, yan2015application}. With the advent of deep learning, neural networks have shown promising capabilities in analyzing historical price series~\cite{chen2018incorporating, ding2020hierarchical, koa2024learning}.

However, stock selection poses greater challenges compared to general time-series tasks. Stock prices exhibit high volatility due to numerous influencing factors in the real market, introducing additional complexities for prediction~\cite{cutler1988moves, fama1995random}. Therefore, recent studies have focused on extracting more robust profit-related information from stock prices by building connections among different stocks~\cite{feng2019temporal}, mining spatio-temporal information~\cite{sawhney2021stock, xia2024ci}, and studying multi-scale patterns~\cite{wang2022adaptive}.

While existing works have highlighted the importance of considering market interactions, pre-training, a widely used representation learning technique in other fields like natural language processing (NLP) and computer vision (CV), has been less explored for stock data. Current pre-training on stock data primarily focuses on two directions: contrastive learning~\cite{hou2021stock} and masked value prediction~\cite{xia2024ci}. However, both approaches have limitations in adapting to stock data. Contrastive learning methods typically have more stringent data requirements, such as minute-level or second-level prices, which are impractical in many situations. The masked value prediction methods, following general time-series pre-training approaches~\cite{zerveas2021transformer}, do not adapt well to stock data, due to the non-stationary nature of stock prices and the complex market dependencies~\cite{wang2021hierarchical}. Therefore, there exists a gap in exploring pre-training methods for stock data to fully extract predictive power from price series. 

\begin{figure}[b]
\centering
\vspace{-0.6cm}
\begin{center}
   \includegraphics[width=0.96\linewidth]{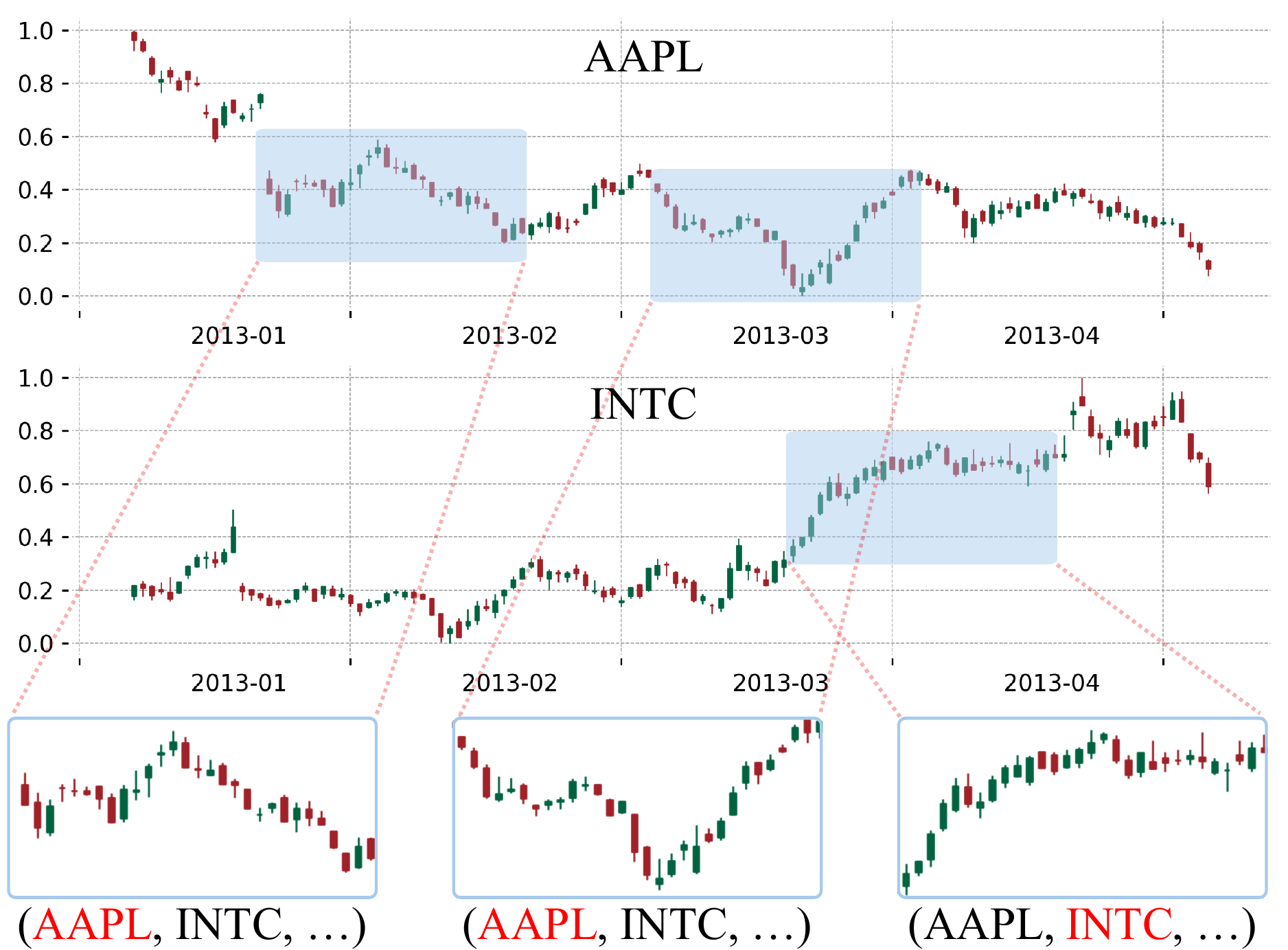}
   \vspace{-0.2cm}
   \captionof{figure}{An example of stock code classification. This task explores whether price series slices from different stocks contain distinguishing features. All prices are normalized to the range of 0 to 1.}
\label{fig:motivation}
\end{center}
\vspace{-0.0cm}
\end{figure}

Designing customized pre-training methods for stock data has the potential to improve stock selection because the unique characteristics of stock data, such as volatility and many other statistical features, are likely underutilized. Since profit is the primary objective of stock prediction tasks, most training objectives are directly related to price changes. However, pre-training research~\cite{devlin2018bert, zhu2020deformable} and exploration of multi-task learning~\cite{zhang2021survey} in other areas have shown that a specific task can benefit from the features learned from other tasks, which cannot be fully used by the task itself. Therefore, exploring other training objectives for stock data is promising to extract more information from stock prices.

In this paper, we propose three specialized pre-training tasks, stock code classification, stock sector classification, and moving average prediction. These tasks require only easily accessible daily price data and basic company information. We implement these tasks using a simple two-layer transformer architecture, which we call the Stock Specialized Pre-trained Transformer (SSPT). Experiments demonstrate that these pre-training tasks effectively capture stock price characteristics and improve stock selection performance, enabling SSPT to outperform existing methods.

The first two tasks, stock code classification and stock sector classification, are designed to identify the stock or sector to which a given price series slice belongs (as illustrated in Figure~\ref{fig:motivation} with an example of stock code classification). While these tasks focus on objectives not directly related to profits, they help capture unique characteristics of price series. Experiments demonstrate that these two tasks effectively capture distinctive price patterns and benefit stock selection, enabling SSPT to outperform existing methods.

The third task, moving average prediction, adapts the widely used masked value prediction idea to stock data. The non-stationary nature of stock prices indicates the presence of random content in price changes, making it impossible to accurately predict each price~\cite{wang1996stock}. Traders use technical indicators like moving average values to obtain relatively stable price features. Inspired by this, we propose predicting the average values from a period after masking some prices, rather than predicting specific masked values.

We evaluate our approach on data from NASDAQ, NYSE, and TOPIX-100 markets over a five-year period, following recent stock selection studies~\cite{feng2019temporal, wang2022adaptive, xia2024ci}. We also test SSPT on FTSE-100 and recent NASDAQ data, to demonstrate our methods can generalize across markets and time periods. Results show that our pre-training methods enhance cumulative investment return ratio (IRR) and Sharpe ratio (SR), outperforming both market benchmarks and existing methods. We provide detailed analyses of pre-training settings and fine-tuning strategies. Additionally, we conduct experiments on simulated data to explore the reasons for the effectiveness of our methods.
In summary, our main contributions are:
\setlength{\parskip}{-5pt}
\begin{enumeratesquish}{0em}{1.0em}
\item[(1)] We propose three customized pre-training tasks for stock data: stock code classification, stock sector classification, and moving average prediction. We demonstrate their effectiveness in enhancing stock selection and provide guidelines for their use.

\item[(2)] We introduce the Stock Specialized Pre-trained Transformer (SSPT), a simple yet effective model based on our pre-training methods. SSPT outperforms market benchmarks and existing methods on IRR and SR across five datasets spanning different markets and time periods.

\item[(3)] Through ablation studies and simulations, we investigate why our pre-training tasks are effective, highlighting underutilized information in stock prices and providing insights into price series analysis.
\end{enumeratesquish}
\setlength{\parskip}{0pt}

\section{Related Work}
\subsection{Stock Prediction and Selection}
Stock prediction has been a longstanding area of research. Early works developed statistical models, such as the ARIMA model, for technical analysis and price prediction~\cite{wang1996stock}. In addition, machine learning models like Hidden Markov Models (HMM) and Support Vector Machines (SVM) have been explored for this purpose~\cite{kavitha2013stock, nayak2015naive}. With the rise of deep learning, neural networks, particularly recurrent neural networks (RNNs) and their variants, have achieved great success in stock forecasting~\cite{qin2017dual, bao2017deep, zhang2017stock}. Later, the transformer structure revolutionized the field of deep learning, and these models have been widely adopted for stock prediction tasks~\cite{wang2022stock, ding2020hierarchical, liu2019transformer}. 

Recently, researchers have recognized the importance of considering market relationships in identifying profitable opportunities, moving beyond treating stocks independently~\cite{chen2018incorporating}. This shift has led to the formulation of the stock selection task, where the goal is to predict multiple stocks collectively and select the most profitable ones. A benchmark dataset was created to facilitate this task~\cite{feng2019temporal}. Graph Neural Networks (GNNs) have been applied from various angles to analyze the spatio-temporal dynamics of stock markets. For instance, some studies have explored hypergraph-based models~\cite{sawhney2021stock}, while others focused on adaptive price patterns~\cite{wang2022adaptive}. These methods have demonstrated effectiveness on large stock markets such as NASDAQ and NYSE. 

However, while existing approaches have extensively explored model architectures to enhance stock analysis, the training objectives and stock feature representations have received limited attention. Most models are trained to minimize movement prediction errors, potentially leading to the underutilization of stock data information. Although features closely related to profitability are well-considered, other features that may seem less directly relevant to profits are likely to be overlooked. These overlooked features may hold potential to improve the understanding of price series and enhance prediction performance.

\begin{figure*}[t]
\centering
\vspace{-0.0cm}
\begin{center}
   \includegraphics[width=0.96\linewidth]{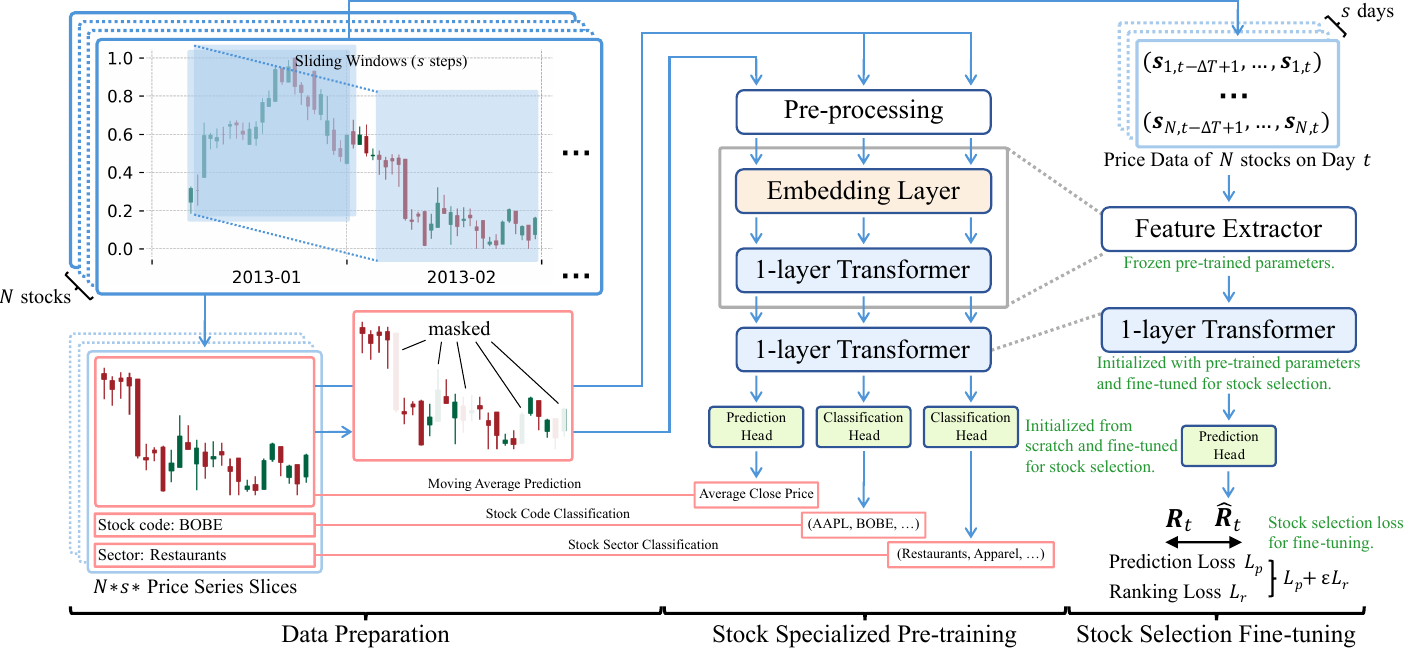}
   \vspace{-0.2cm}
   \captionof{figure}{Overview of the pre-training and fine-tuning procedures for our SSPT model. The frozen parameters can be adjusted, and we conduct experiments to explore the impact of freezing different model components in Section~\ref{sec:independent_pre-training_analysis}.}
\label{fig:model}
\end{center}
\vspace{-0.5cm}
\end{figure*}

\subsection{Neural Networks Pre-training}
Numerous studies have demonstrated the effectiveness of pre-training across various domains, from NLP, CV to Vision-and-Language (VL) tasks~\cite{devlin2018bert, brown2020language, zhu2020deformable}. Pre-training has been shown to benefit downstream tasks by facilitating the learning of high-quality features~\cite{wang2022cmg, hu2019transformation, liu2019roberta}. Inspired by successes in NLP and CV, pre-trained models for time series analysis have also gained attention in recent research~\cite{zerveas2021transformer, zhang2022self, zhou2023one}. These works adapt pre-training methods, such as masked value prediction, which are commonly used in NLP and CV, to time series data for enhanced feature extraction. 

As a specialized type of time series data, stock prices have also been explored using pre-training methods, primarily in two directions: contrastive learning~\cite{hou2021stock} and masked value prediction~\cite{xia2024ci}. The former aims to improve feature representation by analyzing multi-granularity data, while the latter simulates the masked value prediction strategies from other fields to learn contextual information. However, both approaches are largely transplanted from other domains without adequately considering the unique characteristics of stock data. Contrastive learning methods impose stricter requirements on data and continue to focus solely on time series analysis. Meanwhile, masked value prediction methods struggle to adapt effectively to stock data due to the non-stationary nature of stock prices, which makes precise prediction of specific values impractical~\cite{wang2022adaptive}. Therefore, there remains a need for stock data-specific pre-training tasks to improve price feature learning.

\section{Problem Formulation}
\label{sec:problem_formulation}
To ensure a fair comparison with advanced methods, we follow the stock selection formulation presented in previous works~\cite{feng2019temporal, wang2022adaptive, sawhney2021stock, xia2024ci}. Let $\mathcal{S} = \{s_1, s_2, ..., s_N\}$ denotes a set of $N$ stocks. For a given trading day $t$, each stock $\textbf{s}_i$ is associated with $M$ price features over the past $\Delta T$ days, denoted as $\textbf{X}_{i, t} = \{\textbf{x}_{i, t-\Delta T+1}, ..., \textbf{x}_{i, t}\} \in \mathbb{R}^{M\times \Delta T}$. Among the $M$ features of $\textbf{x}_{i, t}$, there exists a closing price $p_{i, t}$, from which we can calculate the 1-day return ratio as $r_{i, t} = \displaystyle\frac{p_{i, t+1} - p_{i, t}}{p_{i, t}}$. The return ratio $r_{i, t}$ serves as our prediction target, and the predicted return ratios for all stocks on day $t$ are denoted by $\hat{\textbf{R}}_t = (\hat{r}_{1, t}, ... \hat{r}_{N, t})$. 
Based on $\hat{\textbf{R}}_t$, we rank the stocks in descending order of return ratios and select the top-ranked stocks for investment on trading day $t$. Our goal is to build a model $f(\cdot\,; w_f)$ with parameters $w_f$ that predicts the return ratios $\hat{r}_{i, t} = f(\textbf{X}_{i, t}; w_f)$ and select the most profitable stocks for investment.

\section{Methodology}
To acquire high-quality stock price representations, we propose three customized pre-training tasks that leverage the unique characteristics of financial market dynamics. Specifically, we design two classification tasks, stock code classification and stock sector classification, which use stock-specific contextual information. Additionally, we introduce a moving average prediction task that considers the non-stationary nature of stock prices. We use a standard two-layer transformer architecture and pre-train the model using these three tasks. Following pre-training, we freeze a subset of the model parameters to retain the acquired knowledge, and fine-tune the remaining parameters for the stock selection task. Figure~\ref{fig:model} provides an overview of our pre-training tasks and the stock selection framework.

\subsection{Stock Code/Sector Classification (SCC/SSC)}
\label{sec:classification_tasks}
Each stock is unique, corresponding to individual companies with distinct backgrounds, environments, business models, and other characteristics that lead to different reactions to market events. While the Efficient Market Hypothesis (EMH)~\cite{fama1970efficient} suggests that a completely efficient market would immediately reflect all such information and leave no room for prediction, practical markets are rarely fully efficient. Numerous studies have demonstrated that stock markets can be predictable to some extent~\cite{christensen2007effect, zhang2016market, luo2023causality}, indicating the existence of exploitable patterns and market inefficiencies that are not yet fully incorporated into prices.

Based on this observation, we hypothesize that specific patterns distinguishing different stocks must exist. Although these patterns may not directly correlate with profits, they could reveal hidden features in price series beneficial for stock prediction. To explore this hypothesis, we design a stock code classification task. As illustrated in Figure~\ref{fig:model}, we segment price series into equal-length slices (matching the look-back period used for subsequent stock selection) and mix slices from different stocks. The model is then trained to identify the source stock of each slice. Using the notation from Section~\ref{sec:problem_formulation} and denoting the additional classification head parameters as $w_{scc}$, we train the model $f$ with the cross-entropy loss for stock code classification, $\mathcal{L}_{scc}$, calculated as:
\begin{equation}
\label{eq:scc}
\begin{aligned}
\mathcal{L}_{scc} = -\log (f(\textbf{X}_{i,t}; w_f, w_{scc})_i).
\end{aligned}
\end{equation}
Note that the value of $i$ is only contained in our annotation $\textbf{X}_{i,t}$, not in the input value itself.

While stocks are unique, they are also interconnected. Market events can affect multiple related stocks simultaneously, and stocks can influence each other. Companies within the same sector often exhibit similar behavior during major market events due to their related businesses. Therefore, we hypothesize that there exist distinct patterns characterizing different sectors and design a stock sector classification task. Similar to the stock code classification, we mix price series slices and train a model to identify the sector of origin for each slice. With $sec(i)$ denoting the sector that stock $s_i$ belongs to and $w_{ssc}$ representing the added classification parameters, we train the model with the loss of stock sector classification $\mathcal{L}_{ssc}$ as:
\begin{equation}
\label{eq:ssc}
\begin{aligned}
\mathcal{L}_{ssc} = -\log (f(\textbf{X}_{i,t}; w_f, w_{ssc})_{sec(i)}).
\end{aligned}
\end{equation}

\begin{figure}[b]
\centering
\vspace{-0.4cm}
\begin{center}
   \includegraphics[width=0.98\linewidth]{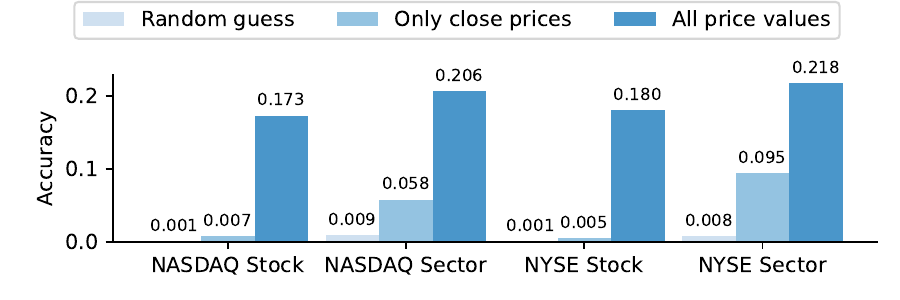}
   \vspace{-0.4cm}
   \captionof{figure}{Classification accuracy comparison for the two pre-training tasks SCC and SSC. The NASDAQ market consists of 1,026 stocks across 112 sectors, and the NYSE market comprises 1,737 stocks in 129 sectors. All results are rounded to three decimal places.}
\label{fig:pretrain}
\end{center}
\vspace{-0.0cm}
\end{figure}

Both pre-training tasks use stock-specific information (stock code and sector) to identify distinguishing patterns in price series. While price series segments often exhibit similarity and complexity that challenge effective classification, our results support the existence of discernible patterns that can differentiate stock codes and sectors. Figure~\ref{fig:pretrain} presents the classification accuracy for both tasks on NASDAQ and NYSE datasets, comparing random guesses, classifications based solely on close prices, and classifications based on all daily price values. Although the highest classification accuracies only reach approximately 0.2, indicating the difficulty of accurately predicting stock source information, these post-training accuracies significantly surpass random guessing. Similar patterns emerge in the other three datasets, as shown in Appendix Section~\ref{sec:appendix_more_classification_results}. This outcome demonstrates the presence of identifiable features within stock price series, which we expect will contribute to improved stock selection performance. 

\subsection{Moving Average Prediction (MAP)}
Stock prices exhibit high volatility and contain significant random noise~\cite{wang1996stock}. Moreover, stock price series are non-stationary, meaning their statistical properties change over time. These characteristics make precise price prediction extremely challenging. When a model achieves very low prediction errors on training and validation sets, it likely indicates overfitting to random fluctuations rather than capturing meaningful patterns.

Despite these challenges, predicting masked prices remains the primary pre-training technique in existing stock prediction methods~\cite{xia2024ci}. Masked value prediction, a widely used pre-training strategy across various fields, has proven effective in predicting masked words in sentences~\cite{devlin2018bert} and masked image patches~\cite{he2022masked}. Consequently, it has been adopted in stock prediction. However, as previously noted, the unique characteristics of price series limit the effectiveness of direct masked value prediction in the stock market context.

To address these limitations, we propose an adaptation of masked value prediction for stock data, inspired by Moving Average (MA) indicators. MA is a common technical indicator in financial analysis that smooths short-term price fluctuations~\cite{murphy1999technical, brock1992simple}. By calculating the average price over a specified period, MA provides a more stable signal compared to raw prices and is widely used by traders for trend analysis. Therefore, since predicting individual prices is unreliable, we propose an alternative approach: predicting the moving average values of a period containing masked values.

As illustrated in Figure~\ref{fig:model}, we calculate the average closing prices within sliding windows. We then mask a portion of the prices within each window and train the model to predict these average prices from the partially masked data. Let $w_{map}$ represents the additional parameters of the prediction head. The model is trained with the loss $\mathcal{L}_{map}$, defined as:
\begin{equation}
\label{eq:map}
\begin{aligned}
\mathcal{L}_{map} = \left(f(\textbf{X}_{i,t}; w_f, w_{map}) - \frac{1}{\Delta T}\sum\nolimits^{\Delta T}_{j=1} p_{i, t-j+1} \right)^2.
\end{aligned}
\end{equation}

This approach mitigates the impact of price volatility and non-stationarity, providing a more robust pre-training task that better aligns with the inherent characteristics of stock data.

\subsection{Pre-training}
We implement our pre-training methods and subsequent stock selection task using a standard two-layer transformer architecture. Apart from the classification and prediction heads necessary for different tasks, we do not add any extra specialized structures. This general model structure, combined with our pre-training methods, forms our Stock Specialized Pre-trained Transformer (SSPT), which proves effective in stock selection.

The basic approach to using these three tasks involves training a model on each task individually, followed by fine-tuning for the stock selection task. However, since the three pre-training tasks are designed to capture different aspects of stock information, it is likely that the model acquires diverse knowledge from each task. Therefore, combining these three tasks during pre-training has the potential to further enhance stock selection performance.

To maximize the benefits of our customized stock pre-training tasks, we explore pre-training the model on multiple tasks simultaneously. Although various combination methods exist, such as feature fusion or sequential pre-training through continual learning, we focus on multi-task pre-training in this paper, following popular approaches in the NLP field~\cite{devlin2018bert}. This involves incorporating multiple task heads into the model and optimizing their corresponding losses simultaneously.

Formally, let $\mathcal{L}_{pt}$ denote the total loss during pre-training, calculated as follows:
\begin{equation}
\label{eq:pre-training}
\begin{aligned}
\mathcal{L}_{pt} = \alpha \mathcal{L}_{scc} + \beta \mathcal{L}_{ssc} + \gamma \mathcal{L}_{map}.
\end{aligned}
\end{equation}
Here, $\alpha$, $\beta$, and $\gamma$ are coefficients that control the relative influence of each corresponding task's loss during pre-training. We conduct experiments with various combinations of these coefficients to explore the balance among the three tasks, as detailed in Section~\ref{sec:combined_pre-training_analysis}.

\subsection{Fine-tuning}
Following pre-training, we adapt the model for stock selection by replacing task-specific output layers with a profit ratio prediction head. In line with established approaches~\cite{feng2019temporal, sawhney2021stock, xia2024ci}, we incorporate both profit ratio regression loss and profit ranking loss during fine-tuning. The stock selection loss $\mathcal{L}_{ft}$ is calculated as:
\begin{equation}
\scriptsize 
\label{eq:finetune}
\begin{aligned}
\mathcal{L}_{ft} = \sum\nolimits^N_{i=1}(\hat{r}_{i, t} - r_{i, t})^2 + \epsilon \sum\nolimits^N_{i=1} \sum\nolimits^N_{j=1} \max(0, -(\hat{r}_{i, t}-\hat{r}_{j, t})(r_{i, t}-r_{j, t})).
\end{aligned}
\end{equation}
where $\epsilon$ is a hyper-parameter balancing the two loss components.

During fine-tuning, the model's parameters are categorized into three groups. The parameters of the initial layers are kept frozen to preserve the pre-trained knowledge, following common practices in pre-training methods~\cite{devlin2018bert, he2022masked}. The intermediate layers, between the frozen layers and the prediction head, inherit their values from the pre-trained model but are fine-tuned for the stock selection task. Finally, the prediction head, which was not involved in the pre-training tasks, is initialized from scratch and trained specifically for stock selection.

This approach, shown in Figure~\ref{fig:model}, enables flexible fine-tuning strategies. As the categorization of parameters can significantly impact performance~\cite{merchant2020happens}, we explore various parameter arrangements to optimize stock selection effectiveness, as detailed in Section~\ref{sec:independent_pre-training_analysis}.

\section{Experiment Settings}
\subsection{Data}
We evaluate our approach using historical stock price data from five markets: NASDAQ (2013-2017), NYSE (2013-2017), FTSE-100 (2013-2017), TOPIX-100 (2016-2020), and NASDAQ-recent (2018-2022). The NASDAQ, NYSE, and TOPIX-100 datasets serve as established benchmarks in stock selection research~\cite{feng2019temporal}, and numerous studies in this field have used them to ensure comparable results~\cite{sawhney2021stock, wang2022adaptive, xia2024ci}. We additionally include FTSE-100 and NASDAQ-recent datasets to evaluate our methods across different markets and time periods. Detailed information of datasets are provided in Appendix Section~\ref{sec:appendix_dataset_statistics}. In line with previous studies, we chronologically partition the data into training (3 years), validation (1 year), and testing (1 year) sets.
The historical data comprises five daily values: open price, high price, low price, close price, and trading volume. Following previous works, we augment these values with 5, 10, 20, and 30-day moving averages. All features undergo min-max normalization based solely on training data statistics.

\subsection{Evaluation Metrics}
We assess model performance using a daily buy-hold-sell trading strategy, measuring the cumulative Investment Return Ratio (IRR) and Sharpe ratio (SR), consistent with standard practice in related literature~\cite{feng2019temporal, sawhney2021stock, wang2022adaptive, xia2024ci}. The strategy involves buying the top $k$ stocks based on model predictions and selling them at the next day's close.

Formally, the IRR on day $t$ is calculated as $\mathrm{IRR}_t = \sum\nolimits_{i \in \hat{\mathcal{S}}_t}r_{i, t}$, where $\hat{\mathcal{S}_t}$ represents the selected stock set on day $t$. The SR measures the risk-adjusted return of a portfolio and is calculated as $\mathrm{SR} = \displaystyle\frac{E(R_p)-R_f}{\mathrm{std}(R_p)}$, where $R_p$ is the profit over the tested period, $E(R_p)$ and $\mathrm{std}(R_p)$ denote the expectation and standard deviation of the profits, and $R_f$ means the risk-free profit. These two metrics are widely used in financial prediction studies~\cite{ye2020reinforcement, yuemei2021predicting, wang2024mana}. Previous research~\cite{feng2019temporal} has explored the impact of varying the value of $k$ and identified $k=5$ as the most representative for performance assessment. Most subsequent works report results based on this $k=5$ setting, and we also adhere to it.

\subsection{Baseline Methods}
We conduct a comprehensive comparisons with various baseline methods, including recent SOTA methods. The baselines are categorized into four groups: classification (CLF) models, regression (REG) models, reinforcement learning (RL) methods, and ranking (RAN) approaches. Table~\ref{tab:baseline_comparison} provides brief descriptions of these methods.

\subsection{Implementation Details}
Our SSPT model uses a standard two-layer transformer architecture. The embedding layer consists of a feature embedding layer and a positional embedding layer. Both the task heads for pre-training and the stock selection prediction head are implemented as fully connected layers with corresponding output dimensions. We set the hidden size of attention vectors to 32 and use 4 attention heads. For other intermediate features, we use a hidden size of 128. Following previous studies, we select the look-back length from the set $\{16, 32\}$. The model is trained using the Adam optimizer, with the learning rate chosen from $\{10^{-3},  10^{-4},  10^{-5}\}$. The hyper-parameter $\epsilon$, which balances the prediction and ranking losses, is selected from $\{1, 5, 10\}$. To facilitate ranking loss computation, the batch size is set equal to the number of stocks $N$. Both pre-training and fine-tuning phases are limited to 100 epochs, with the optimal epoch and hyper-parameter combination determined based on validation set performance. The model training process, including both pre-training and fine-tuning, is conducted using the training and validation sets, while the reported results reflect the stock selection performance on the testing split.

\section{Results and Discussion}
\label{section:results}
Our analysis aims to provide comprehensive insights and practical guidelines for effectively implementing our customized stock pre-training tasks, rather than merely demonstrating performance improvements. Therefore, we structure our investigation around three research questions (RQs):
\setlength{\parskip}{-5pt}
\begin{enumeratesquish}{0em}{0.0em}
\item[\textbf{RQ1}] How can we maximize the benefits of stock pre-training for the subsequent stock selection task?
\item[\textbf{RQ2}] How does our stock selection framework SSPT compare to existing approaches?
\item[\textbf{RQ3}] Why do these pre-training tasks improve stock selection performance?
\end{enumeratesquish}
\setlength{\parskip}{0pt}

Section~\ref{sec:independent_pre-training_analysis} and Section~\ref{sec:combined_pre-training_analysis} address \textbf{RQ1} from two perspectives: optimizing the performance of individual pre-training tasks and effectively combining multiple pre-training tasks. Section~\ref{sec:stock_selection_results} tackles \textbf{RQ2} through comprehensive baseline comparisons. Section~\ref{sec:simulation_results} answers \textbf{RQ3} through experiments on controlled simulated data. 

While we present results across five datasets for baseline comparisons in Section~\ref{sec:stock_selection_results}, our detailed analysis in other sections primarily focuses on the NASDAQ dataset. We present SR results when analyzing pre-training tasks, as the analysis based on SR or IRR can reflect similar conclusions. However, SR is considered the more important metric as it incorporates not only returns but also risks, which are crucial factors in market investment~\cite{wang2024mana}. We also include the corresponding IRR results and analysis in Appendix Section~\ref{sec:appendix_IRR_results}.


\subsection{Analysis of Individual Pre-training Tasks}
\label{sec:independent_pre-training_analysis}
Our analysis of individual pre-training tasks reveals two critical hyper-parameters: input features and learning rate. As presented in Figure~\ref{fig:pretrain} (Section~\ref{sec:classification_tasks}), incorporating all daily price values as input features substantially improves the accuracy for the two classification tasks, compared to using only close prices. The MAP task exhibits a similar trend, with the optimized mean squared error (MSE) decreasing from $5.5\times10^{-6}$ when using only closing prices to $2.3\times10^{-6}$ when using all price values. This consistent finding across tasks highlights the importance of using comprehensive price information during pre-training.

\begin{figure}[b]
\centering
\vspace{-0.4cm}
\begin{center}
   \includegraphics[width=0.98\linewidth]{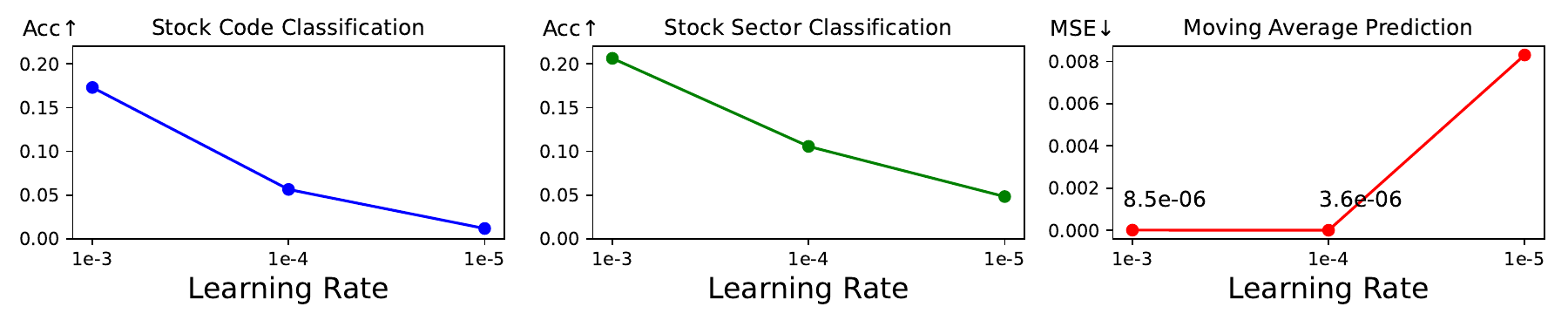}
   \vspace{-0.4cm}
   \captionof{figure}{Evaluation results of the three pre-training tasks at varying learning rates. Note that the metrics differ among tasks: higher accuracy is better, whereas lower MSE is better.}
\label{fig:lr_comparison}
\end{center}
\vspace{-0.0cm}
\end{figure}

However, the optimal learning rates vary across tasks. As shown in Figure~\ref{fig:lr_comparison}, the pre-training tasks demonstrate sensitivity to the learning rate, with the classification tasks (SCC and SSC) performing best with a higer learning rate ($10^{-3}$), while the value prediction task (MAP) requires a lower rate ($10^{-4}$).

After optimizing the pre-trained models, we explore effective fine-tuning strategies for stock selection. A crucial aspect during fine-tuning is the management of model parameters. While the added task-specific prediction head is a confirmed component, the frozen and fine-tuned parameters can be flexibly adjusted. Although Figure~\ref{fig:model} shows an example of freezing the entire feature extractor, this may not be optimal, as research in NLP has shown that parameter freezing strategies significantly impact performance~\cite{merchant2020happens}.

We evaluate four fine-tuning strategies: (1) freezing no parameters, (2) freezing only the embedding layers, (3) freezing the embedding and attention layers, and (4) freezing the entire feature extractor. Note that freezing no parameters is equivalent to using the pre-trained parameters to initialize the model for stock selection. Additionally, the third strategy differs from the fourth due to an additional feed-forward layer after the attention layer not shown in Figure~\ref{fig:model}.

\begin{figure}[t]
\centering
\begin{center}
   \includegraphics[width=1.0\linewidth]{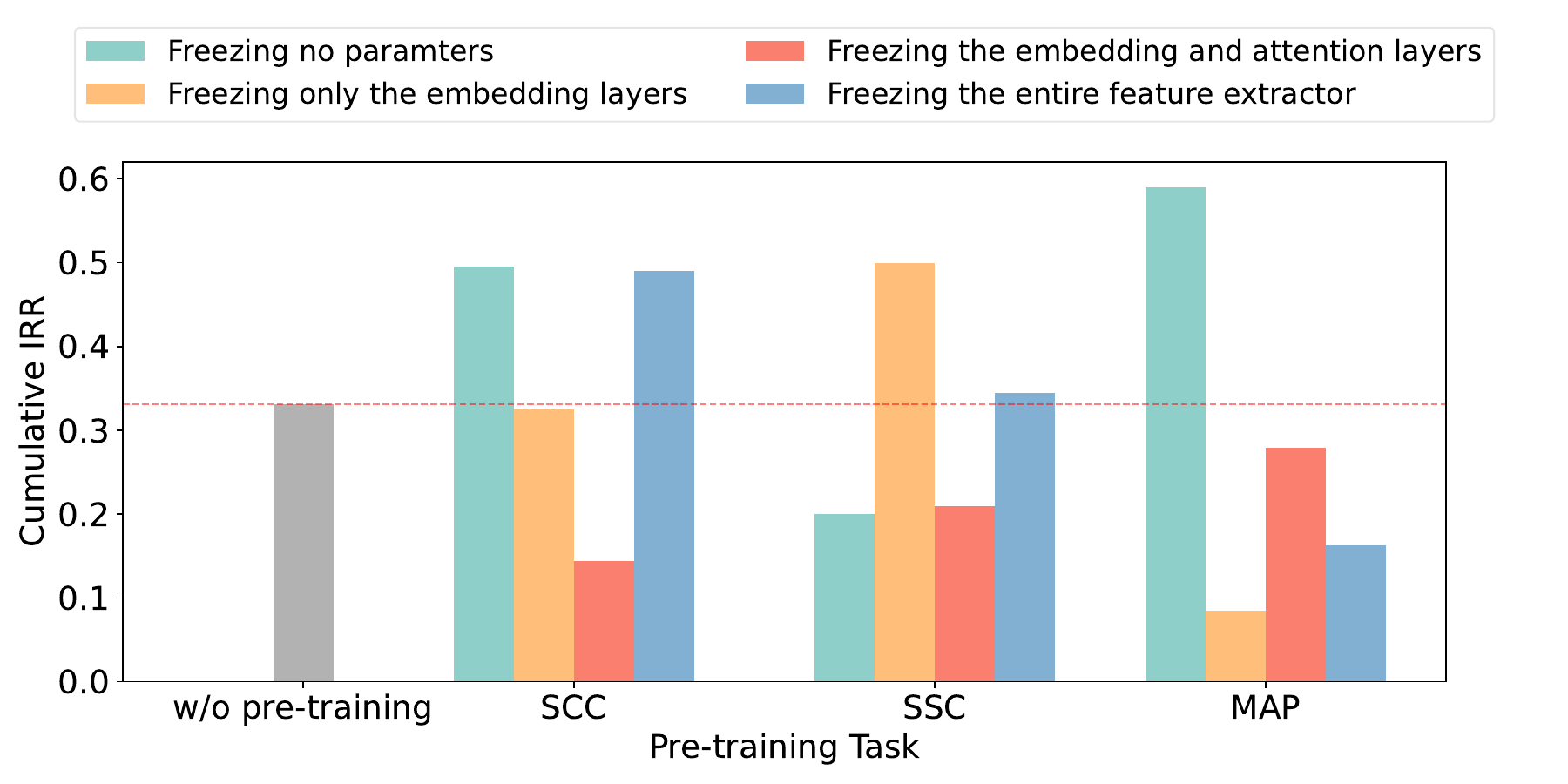}
   \vspace{-0.8cm}
   \captionof{figure}{The Sharpe ratio results of the stock selection task with different fine-tuning strategies. The group `w/o pre-training' presents the result of directly training the model on the stock selection task without any pre-training.}
\label{fig:ft_comparison}
\end{center}
\vspace{-0.6cm}
\end{figure}

The results of these strategies are presented in Figure~\ref{fig:ft_comparison}, where we include results of direct training without pre-training as a control group. The three pre-training tasks show distinct responses to fine-tuning strategies. SCC and SSC generally improve stock selection performance across most strategies, with SCC performing best when no parameters are frozen, and SSC favoring freezing only the embedding layer. However, the MAP task shows high sensitivity to the fine-tuning strategy. It can significantly enhance performance when no parameters are frozen but harm the subsequent task when using other strategies. This suggests that MAP provides effective model initialization for stock selection but cannot directly transfer its learned knowledge. 
This phenomenon can be attributed to the structural similarity between MAP and stock selection tasks. Since both tasks involve price-based value regression, they likely train the model to extract similar features for different targets. Consequently, MAP can establish a favorable initial model for stock selection, but its learned features cannot be directly transferred to the target task.

Furthermore, the comparison with the control group serves as an ablation study, confirming that our three pre-training tasks are indeed beneficial for the subsequent stock selection task.

In summary, through the analysis of individual pre-training tasks, we identify the most influential factors: pre-training feature selection, learning rate, and fine-tuning strategy. Additionally, an ablation study comparing traditional masked value prediction with our proposed MAP approach, along with an analysis of different mask rates, is presented in Appendix Section~\ref{sec:appendix_map_ablation}. Based on these results, we recommend the following guidelines: (1) include the complete set of price features during pre-training, (2) use a learning rate of $10^{-3}$ for SCC and SSC but $10^{-4}$ for MAP during pre-training, and (3) during fine-tuning, do not freeze any parameters for SCC and MAP but freeze the embedding layer for SSC.

\subsection{Analysis of Combined Pre-training Tasks}
\label{sec:combined_pre-training_analysis}
After analyzing the three individual pre-training tasks, we explore potential performance improvements through task combinations.
During combined pre-training, in addition to the input features and learning rate, the coefficients of the loss terms ($\alpha, \beta, \gamma$) in Equation~\ref{eq:pre-training} also impact performance. Table~\ref{tab:multi-task} presents the results for balanced multi-task pre-training ($\alpha=\beta=\gamma=1$). Notably, each task's performance degrades when trained simultaneously with others, indicating conflicts between different training objectives.

\begin{table}[t]
\vspace{-0.0cm}
\renewcommand\arraystretch{1}
\captionof{table}{Evaluation results of pre-training with task combinations. The first row presents the results of pre-training with individual tasks. The following four rows show the results when the loss coefficients of the three tasks are equal, meaning $\alpha=\beta=\gamma=1$ in Equation~\ref{eq:pre-training}.}
\vspace{-0.4cm}
\begin{center}
\small
\begin{tabular}[b]{l c c c}
\toprule
Pre-training Tasks & Acc of SCC & Acc of SSC & MSE of MAP \\
\cmidrule(r){1-4}
SCC / SSC / MAP & 0.173 & 0.206 & 8.5e-6 \\
SCC + SSC & 0.129 & 0.124 & \textbackslash \\
SCC + MAP & 0.088 & \textbackslash & 2.5e-3 \\
SSC + MAP & \textbackslash & 0.097 & 7.3e-4 \\
SCC + SSC + MAP & 0.069 & 0.069 & 3.1e-3 \\
\bottomrule
\end{tabular}
\end{center}
\label{tab:multi-task}
\vspace{-0.4cm}
\end{table}

Through extensive experimentation, we observe several key patterns. The conclusion regarding input features follows the previous section: using all price values consistently yields better results. The learning rate is confirmed to be $10^{-3}$ for all combinations, as when combined with other tasks, MAP's MSE results are not sensitive to the learning rate, remaining around the values in Table~\ref{tab:multi-task}. Regarding loss coefficients, increasing $\alpha$ to 5 marginally improves SCC accuracy without compromising SSC performance. Increasing $\gamma$ reduces the MSE of MAP but significantly harms the classification tasks. Given the uncertain relationship between pre-training balance and stock selection performance, we explore various coefficient combinations and select the best model based on validation set results for further investigation into how combined pre-training influences subsequent stock selection.

\begin{figure}[b]
\vspace{-0.6cm}
\centering
\begin{center}
   \includegraphics[width=0.98\linewidth]{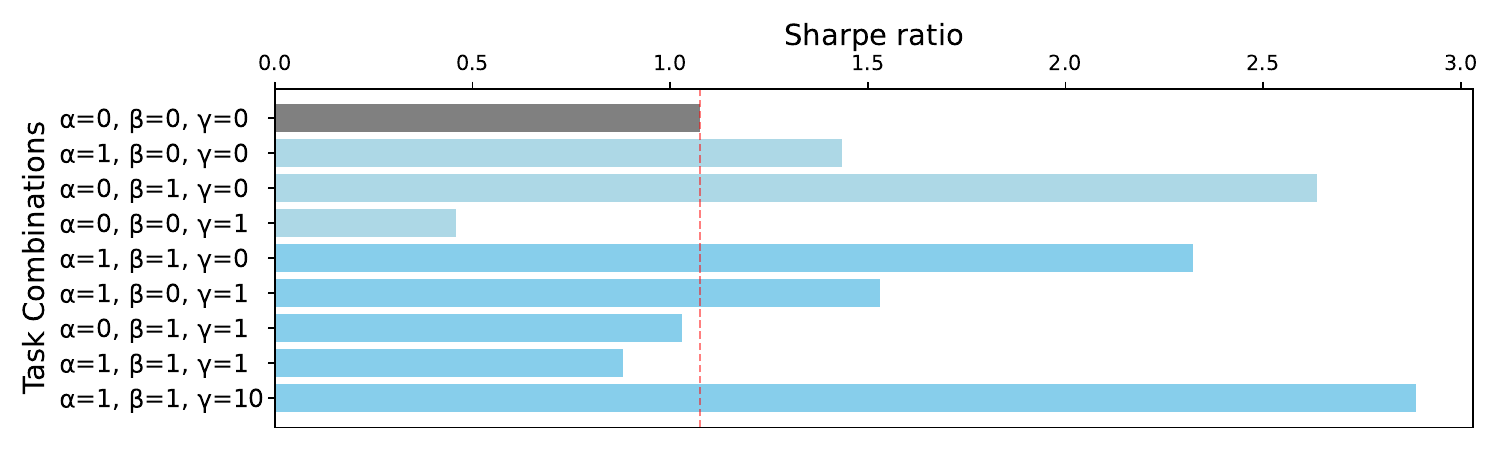}
   \vspace{-0.2cm}
   \captionof{figure}{Sharpe ratio results on the NASDAQ market for the stock selection task using pre-trained models with different combinations of pre-training tasks. The values of $\alpha, \beta, \gamma$ represent the loss coefficients for the SCC, SSC, and MAP tasks in the combined pre-training loss (Equation~\ref{eq:pre-training}). When a coefficient is 0, the corresponding task is excluded from pre-training. The first row shows the baseline result without pre-training. The next three rows display results when pre-training on individual tasks (using the fine-tuning strategy of freezing only the embedding layers). Subsequent rows present results from combinations of pre-training tasks.}
\label{fig:ft_combination}
\end{center}
\vspace{-0.0cm}
\end{figure}

Figure~\ref{fig:ft_combination} presents the Sharpe ratio results for several representative pre-training combinations. Given MAP's significant performance degradation (higher MSE) in combined settings, we prioritize SCC and SSC in determining the fine-tuning strategy, freezing only the embedding layers. We observe that the combination of SCC and SSC maintains good performance. However, the combinations including the MAP task are different. Although combined training presents improved SR results compared with using MAP individually, two out of the three combinations including MAP are worse than not using pre-training, suggesting that MAP is not an optimal component for combined pre-training. These findings are consistent with our previous conclusions, as MAP favors different settings from the other two tasks, and the MAP task is negatively affected significantly when combined with other tasks in pre-training. 

However, the last line in Figure~\ref{fig:ft_combination} presents an interesting case: combining all three tasks with increased MAP coefficient ($\gamma=10$), yields a Sharpe ratio exceeding the SCC+SSC combination. This indicates that MAP includes knowledge that the other two tasks cannot capture. However, as this coefficient combination highly relies on validation selection, and other coefficients with higher $\gamma$ values, like $\gamma=5$, do not consistently present the same trend, including MAP in the combined pre-training tasks still indicates potential robustness issues.

In summary, we find that the combination of SCC and SSC pre-training can reliably improve the stock selection results. However, the inclusion of MAP is difficult to determine as its performance is sensitive to the loss coefficient settings. Therefore, a relatively robust recipe for pre-training task combination is to combine SCC and SSC with balanced coefficients ($\alpha=\beta=1$).

\subsection{Comparison with Existing Methods}
\label{sec:stock_selection_results}

\begin{table*}[ht]
\vspace{-0.0cm}
\captionof{table}{Comparison of stock selection performance on NASDAQ, NYSE, and TOPIX-100 markets. Our SSPT models using individual pre-training tasks (SSPT-ind) and combined pre-training tasks (SSPT-comb) are compared with  various existing methods. Bold and underline values indicate the best and second-best results, respectively.}
\vspace{-0.2cm}
\begin{center}
\begin{tabular}[b]{l l l@{\hspace{4pt}} c c c c c c}
\toprule

 & \multirow{2}{*}{Model} & \multirow{2}{*}{Description} & \multicolumn{2}{c}{NASDAQ} & \multicolumn{2}{c}{NYSE} & \multicolumn{2}{c}{TOPIX-100} \\
 \cmidrule(r){4-5} \cmidrule(r){6-7} \cmidrule(r){8-9} 
 & & & IRR & SR & IRR & SR & IRR & SR \\
\hline
& Market & Select all stocks, showing basic market
performance. & 0.15 & 1.53 & 0.10 & 1.49 & 0.02 & 0.19 \\
\hline
\multirow{7}{*}{CLF}
 & ARIMA~\cite{wang1996stock} (1996) 
 & RNN with features from ARIMA analysis. 
 &  0.10 & 0.55 & 0.10 & 0.33 & 0.13 & 0.47 \\
 & Adv-ALSTM~\cite{feng2019enhancing} (2019) 
 & Adversarial training for better prediction generalization. 
 & 0.23 & 0.97 & 0.14 & 0.81 & 0.43 & 1.10 \\
 & HGCluster~\cite{luo2014stock} (2014) 
 & Use a hyper-graph model to predict the stock trends. 
 & 0.10 & 0.06 & 0.11 & 0.10 & 0.10 & 0.20 \\
 & HATS~\cite{kim2019hats} (2019) 
 & A hierarchical attention network using relational data.
 & 0.15 & 0.80 & 0.12 & 0.73 & 0.31 & 0.96 \\
 & HMG-TF~\cite{ding2020hierarchical} (2020) 
 & Use a Multi-Scale Gaussian Prior to improve model. 
 & 0.19 & 0.83 & 0.13 & 0.75 & 0.33 & 1.05 \\
 & LSTM-RGCN~\cite{li2021modeling} (2021) 
 & Model stock connections by their correlation matrix. 
 & 0.13 & 0.75 & 0.10 & 0.70 & 0.28 & 0.90 \\
 & DTML~\cite{yoo2021accurate} (2021) 
 & Learn the correlations between stocks for prediction. 
 & 0.41 & 1.35 & 0.45 & 1.17 & 0.35 & 1.07 \\
 & HATR~\cite{wang2022hatr} (2022) 
 & Grasp multi-scale transition regularities of stocks. 
 & 0.31 & 0.92 & 0.14 & 0.76 & 0.36 & 0.98 \\
\hline
\multirow{2}{*}{REG}
 & SFM~\cite{zhang2017stock} (2017) 
 & A State Frequency Memory model. 
 & 0.09 & 0.16 & 0.11 & 0.19 & 0.07 & 0.08 \\
 & DA-RNN~\cite{qin2017dual} (2017) 
 & A dual-stage attention-based model. 
 & 0.14 & 0.71 & 0.13 & 0.66 & 0.25 & 0.86 \\
 & TimeMixer~\cite{wang2024timemixer} (2024) 
 & Analyze temporal variations by multiscale-mixing. 
 & 0.42 & 1.64 & 0.23 & 1.23 & 0.30 & 0.93 \\
 & StockMixer~\cite{fan2024stockmixer} (2024) 
 & Use influences between stock and market. 
 & 0.20 & 1.40 & 0.54 & 1.57 & 0.33 & 1.12 \\
 & Master~\cite{li2024master} (2024) 
 & Use market information for automatic feature selection. 
 & 0.24 & 1.20 & 0.23 & 1.27 & 0.25 & 0.95 \\
\hline
\multirow{3}{*}{RL}
 & DQN~\cite{carta2021multi} (2021) 
 & An ensemble of deep Q-learning agents. 
 & 0.20 & 0.93 & 0.12 & 0.72 & 0.31 & 1.08 \\
 & iRDPG~\cite{liu2020adaptive} (2020) 
 & Deep reinforcement learning and imitation learning. 
 & 0.28 & 1.32 & 0.18 & 0.85 & 0.55 & 1.10 \\
 & RAT~\cite{xu2021relation} (2021) 
 & A relation-aware Transformer with RL. 
 & 0.40 & 1.37 & 0.22 & 1.03 & \underline{0.64} & 1.20 \\
\hline
\multirow{6}{*}{RAN}
 & RSR-I~\cite{feng2019temporal} (2019) 
 & A temporal GCN capturing stock relations. 
 & 0.39 & 1.34 & 0.21 & 0.95 & 0.53 & 1.08 \\
 & STHAN-SR~\cite{sawhney2021stock} (2021) 
 & A neural hyper-graph architecture for stock selection. 
 & 0.44 & 1.42 & 0.33 & 1.12 & 0.62 & 1.19 \\
 & MTSR~\cite{ma2022stock} (2022) 
 & Stock ranking with multi-task learning. 
 & 0.30 & 1.58 & 0.57 & 1.36 & 0.33 & 1.03 \\
 & ALSP-TF~\cite{wang2022adaptive} (2022) 
 & An adaptive long-short patter transformer. 
 & 0.53 & 1.55 & 0.41 & 1.24 & \textbf{0.71} & \underline{1.27} \\
 & TSPRank~\cite{li2024tsprank} (2024) 
 & A hybrid pairwise-listwise ranking method.
 & 0.29 & 1.43 & 0.28 & 1.74 & 0.35 & 1.11 \\
 & CI-STHPAN~\cite{xia2024ci} (2024) 
 & A spatio-temporal hyper-graph model with pre-training. 
 & 0.66 & 2.01 & \textbf{0.79} & \underline{2.14} & 0.28 & 0.91 \\
 & SSPT-ind (ours) 
 & Models based on our individual stock pre-training tasks. 
 & \underline{0.74} & \textbf{2.32} & 0.41 & 2.11 & 0.51 & \textbf{1.33} \\
 & SSPT-comb (ours) 
 & Models based on our combined stock pre-training tasks. 
 & \textbf{0.82} & \underline{2.25} & \underline{0.60} & \textbf{2.35} & 0.43 & 1.21 \\

\bottomrule
\end{tabular}
\end{center}
\label{tab:baseline_comparison}
\vspace{-0.4cm}
\end{table*}

After analyzing the effectiveness and usage of our pre-training tasks, we now demonstrate SSPT's superior performance compared to existing methods and market benchmarks. In addition to various baselines for stock selection, we include a baseline strategy of selecting all available stocks daily to establish basic market performance. Trading strategies that outperform this market baseline are considered as effectively beating the market.

Note that while previous sections' results were based on models selected from specific hyper-parameter setting groups to facilitate detailed analysis, our comparative evaluation uses models selected from a broader range of settings based on validation results, ensuring fair comparison. Consequently, some results here may differ from those previously reported.

Table~\ref{tab:baseline_comparison} compares our methods with various baselines across three markets: NASDAQ, NYSE, and TOPIX-100. These widely-studied datasets enable comprehensive performance comparison. We evaluate our method in two ways: using only individual pre-training tasks (SSPT-ind) and using combined pre-training tasks (SSPT-comb). Both our models consistently outperform the market and achieve the best SR performance, with competitive IRR results across all three datasets.

Table~\ref{tab:more_results} extends our analysis to FTSE-100 and NASDAQ-recent datasets. Due to implementation constraints, such as unavailable codes or extra required information for other methods, we compare these results only against market performance. Having demonstrated SSPT's superiority against various methods in Table~\ref{tab:baseline_comparison}, we focus on showing consistent market outperformance across more exchanges and time periods in Table~\ref{tab:more_results}. Notably, SSPT maintains strong performance even on the NASDAQ-recent dataset, which includes the challenging COVID-19 period.

These comprehensive comparisons demonstrate that SSPT's price series pre-training tasks extract more beneficial knowledge for stock selection than the spatial or temporal information captured by other methods. Using a standard transformer structure, which is widely used as the backbone for the listed baselines, SSPT's SOTA results highlight the effectiveness of our customized stock pre-training approach.

\begin{table}[t]
\vspace{0.1cm}
\renewcommand\arraystretch{1}
\captionof{table}{Comparison of market performance and our SSPT method on FTSE-100 and recent NASDAQ markets.}
\vspace{-0.2cm}
\begin{center}
\begin{tabular}[b]{l p{1.2cm}<{\centering} p{1.2cm}<{\centering} p{1.2cm}<{\centering} p{1.2cm}<{\centering}}
\toprule
 \multirow{2}{*}{Model} & \multicolumn{2}{c}{FTSE-100} & \multicolumn{2}{c}{NASDAQ-recent} \\
 \cmidrule(r){2-3} \cmidrule(r){4-5} 
 & IRR & SR & IRR & SR \\
\hline
Market & 0.17 & 2.08 & 0.16 & 0.97  \\
SSPT-ind & 0.38 & 2.51 & 0.59 & 1.23 \\
SSPT-comb & 0.33 & 2.30 & 0.48 & 1.33 \\
\bottomrule
\end{tabular}
\end{center}
\label{tab:more_results}
\vspace{-0.7cm}
\end{table}

\subsection{Task Analysis on Simulated Data}
\label{sec:simulation_results}
Having demonstrated our pre-training tasks can benefit stock selection, we investigate the underlying mechanisms of their effectiveness. We hypothesize that these tasks enhance stock selection by extracting distinguishing features from price series data. To investigate this hypothesis, we analyze the classification tasks through controlled simulation experiments.

\begin{figure}[b]
\centering
\vspace{-0.4cm}
\begin{center}
   \includegraphics[width=0.98\linewidth]{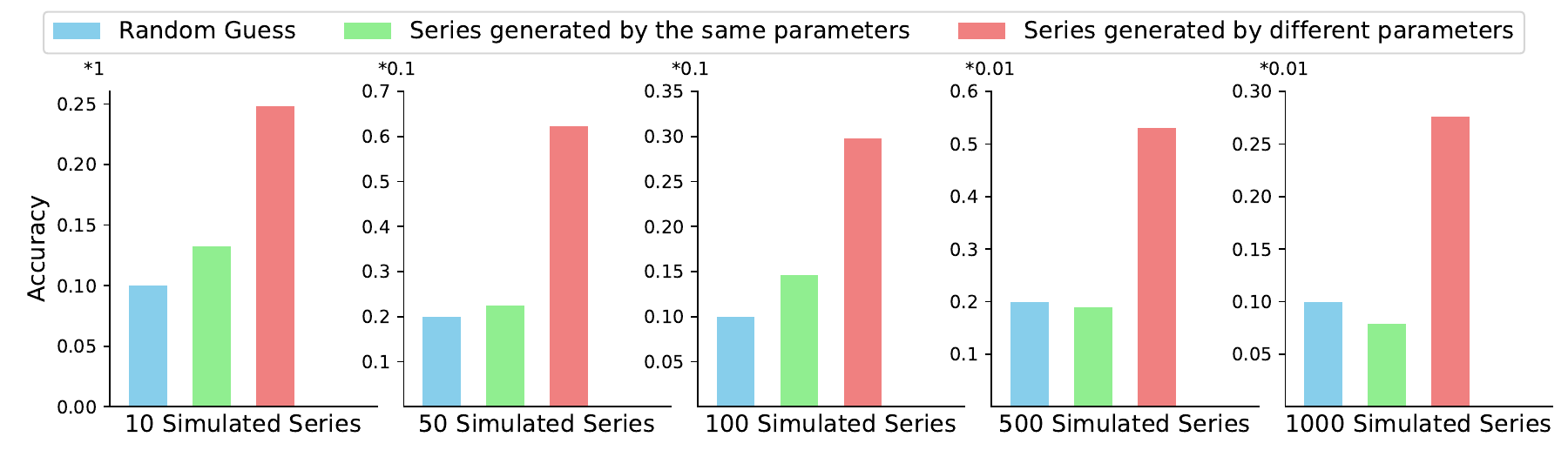}
   \vspace{-0.4cm}
   \captionof{figure}{Classification results on 10, 50, 100, 500, and 1000 simulated series. Each accuracy value is averaged over 5 simulation runs. The y-axis scales are independent for each figure, with the scale factors indicated.}
\label{fig:simulated_data}
\end{center}
\vspace{-0.0cm}
\end{figure}

We use the Wiener Process, a continuous-time stochastic process widely used in finance to model the random behavior of asset prices, to generate simulated stock price series~\cite{shreve2005stochastic, bjork2009arbitrage, hull2016options}. The simulation process is formulated as:
\begin{equation}
\label{eq:witner_process}
\begin{aligned}
S(t + \Delta t) = S(t) * \exp (( \mu - \sigma^2 / 2 ) \Delta t + \sigma \sqrt{\Delta t} Z_t ).
\end{aligned}
\end{equation}
Here, $\Delta t$ represents a small time increment, $Z_t$ is a random variable sampled from a standard normal distribution, $\mu$ and $\sigma$ are the simulation parameters representing the expected return and volatility of the stock. To generate a price series, we need to specify the initial value $S(0)$, the time horizon $t$, and the parameters $\mu$ and $\sigma$.

Following our SCC and SSC task designs, we simulate $N$ price series, segment them into equal-length slices, and train a classification model to identify source series for given slices. We maintain constant $S(0)$ and $t$ while comparing two scenarios: (1) simulating $N$ series with identical random $\mu$ and $\sigma$, and (2) simulating $N$ series with different random $\mu$ and $\sigma$. Figure~\ref{fig:simulated_data} shows classification accuracy of the two scenarios across different values of $N$.

We observe that the classification results for series generated with identical parameters are very close to random guessing, while the accuracy for series generated with different parameters is much higher. This indicates that classification models primarily rely on the different statistical properties of the original series to differentiate series slices. This conclusion supports our hypothesis that the SCC and SSC tasks can extract information related to the statistical characteristics that distinguish various stocks, thereby providing more informative features for the stock selection task. 

While practical stock price series are much more complex and contain more latent statistical features than our simulated data, our results indicate that the pre-training tasks function by extracting distinguishing information.
Additional simulation results and analysis in Appendix Section~\ref{sec:appendix_simulation} further demonstrate the pre-training tasks' sensitivity to price series' statistical features, strengthening our hypothesis.

\section{Conclusions}
This paper introduces three novel pre-training tasks specifically designed for stock price data. Through extensive experiments, we demonstrate the effectiveness of these customized pre-training tasks in enhancing stock selection. Our Stock Specialized Pre-trained Transformer (SSPT) framework, built upon these pre-training methods and a standard transformer architecture, outperforms the market and existing methods on five datasets. We conduct a comprehensive analysis to provide practical guidance on optimally using our pre-training tasks. This includes identifying the most influential factors and the best fine-tuning strategies for individual tasks as well as their combinations. Furthermore, through experiments on simulated data with controlled parameters, we explore the underlying reasons for the effectiveness of our pre-training methods. In summary, this work introduces an effective stock selection framework, provides practical implementation guidelines, and offers insights into stock price series analysis. The demonstrated success of our specialized pre-training approach suggests promising directions for future research in financial market analysis and prediction.

\section*{Acknowledgments}
This work was supported by the UKRI Centre for Doctoral Training (CDT) in Natural Language Processing through UKRI grant EP/S022481/1. We would like to thank Chang Luo, Waylon Li, and the anonymous reviewers for their valuable feedback.

\bibliographystyle{ACM-Reference-Format}
\balance
\bibliography{mypaper}

\appendix

\section{Dataset Introduction}
\label{sec:appendix_dataset_statistics}
Our experiments use five datasets spanning major stock markets across the US, UK, and Japan over different time periods: NASDAQ (2013-2017), NYSE (2013-2017), FTSE-100 (2013-2017), TOPIX-100 (2016-2020), and NASDAQ-recent (2018-2022). NASDAQ, NYSE, and TOPIX-100 serve as established benchmark datasets in stock selection research~\cite{feng2019temporal,wang2022adaptive}, while FTSE-100 and NASDAQ-recent were collected from Yahoo Finance (https://finance.yahoo.com). Sectors are assumed to be constant in each dataset. They can be updated periodically, as we do with NASDAQ and NASDAQ-recent. Additionally, sector shifts are rare and minor for large companies, thus affecting pre-training marginally.

\noindent\textbf{NASDAQ}~\cite{feng2019temporal} is a highly volatile US exchange, comprising 1,026 stocks from 112 sectors, drawn from S\&P 500 and NASDAQ Composite Indexes.

\noindent\textbf{NYSE}~\cite{feng2019temporal} is the world's largest stock exchange by market capitalization, offers relatively stable market conditions compared to NASDAQ. The dataset includes 1,737 stocks across 129 sectors.

\noindent\textbf{TOPIX-100}~\cite{li2021modeling} is a smaller market contrasting with US markets. It represents Japan's major market index, featuring 95 stocks from 10 sectors, focusing on the largest market capitalizations in the Tokyo Stock Exchange.

\noindent\textbf{FTSE-100} is the UK's best-known stock market index, representing the largest market capitalizations on the London Stock Exchange. This dataset includes 87 stocks from 11 sectors.

\noindent\textbf{NASDAQ2} tracks the same companies as the NASDAQ datast but during 2018-2022, a recent period including the challenging COVID time. This dataset includes 718 stocks across 106 sectors.

These datasets provide comprehensive coverage of diverse market conditions, including both established American markets (widely considered most efficient) and representative markets from other regions. The datasets span both growth periods and volatile years, ensuring robust and generalizable evaluation of our methods.

\section{More Results of Classification Tasks}
\label{sec:appendix_more_classification_results}

\begin{figure}[t]
\centering
\vspace{-0.0cm}
\begin{center}
   \includegraphics[width=1.0\linewidth]{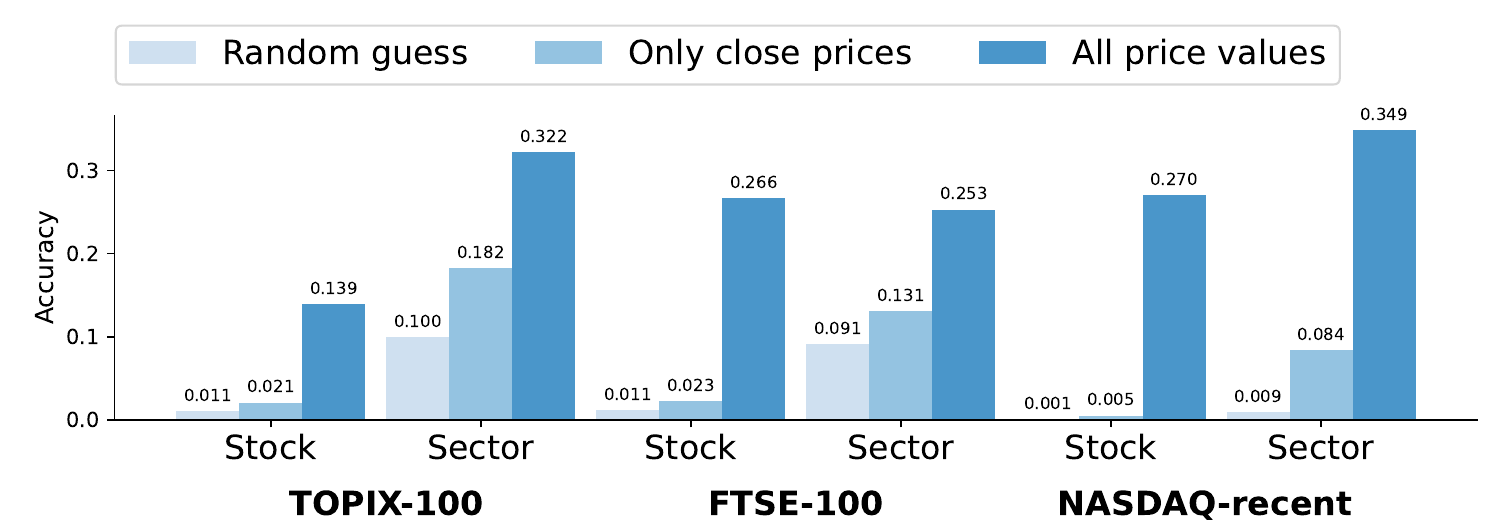}
   \vspace{-0.6cm}
   \captionof{figure}{Classification accuracy comparison for the two pre-training tasks SCC and SSC on TOPIX-100, FSTE-100, and NASDAQ-recent datasets. All results are rounded to three decimal places.}
\label{fig:pretrain2}
\end{center}
\vspace{-0.4cm}
\end{figure}

In Section~\ref{sec:classification_tasks}, we presented classification accuracy results for SCC and SSC on the two largest datasets, NASDAQ and NYSE. Figure~\ref{fig:pretrain2} extends this analysis to the TOPIX-100, FTSE-100, and NASDAQ-recent datasets. The results demonstrate consistent patterns across all datasets: classification accuracy significantly exceeds random guessing, and using comprehensive price values as inputs consistently yields optimal results. These findings reinforce our conclusions from Section~\ref{sec:classification_tasks}: price series contain distinctive information that enables effective classification, and using all price values maximizes feature quality.

\begin{table}[b]
\vspace{-0.2cm}
\renewcommand\arraystretch{1}
\captionof{table}{Comparison between traditional masked value prediction (MVP) and our proposed MAP method on pre-training and downstream stock selection performance.}
\vspace{-0.4cm}
\begin{center}
\begin{tabular}[b]{l c c c c c}
\toprule
 \multirow{3}{*}{Method} & \multicolumn{3}{c}{Pre-training} & \multicolumn{2}{c}{Stock Selection} \\
 \cmidrule(r){2-4} \cmidrule(r){5-6} 
 & MSE & MSE & MSE & IRR & SR \\
 & (lr=1e-3) & (lr=1e-4) & (lr=1e-5) & \multicolumn{2}{c}{(NASDAQ)} \\
\hline
MVP & 7.6e-5 & 9.5e-6 & 1.5e-4 & 0.53 & 1.54 \\
MAP & 8.5e-6 & 3.6e-6 & 8.3e-3 & 0.59 & 1.97 \\
\bottomrule
\end{tabular}
\end{center}
\label{tab:ablation_map}
\vspace{-0.0cm}
\end{table}

\section{Further Analysis of the MAP Task}
\label{sec:appendix_map_ablation}
Since our MAP task is motivated by the traditional masked value prediction (MVP) approach and introduces additional hyper-parameters such as the mask rate, we conduct further experimental analysis to better understand its behavior.

First, we compare our MAP method with the traditional MVP approach to validate the effectiveness of our design tailored for stock data. Both pre-training tasks are conducted on the NASDAQ dataset, followed by a downstream stock selection task. We report MSE during pre-training and IRR and SR during the stock selection phase. The results are summarized in Table~\ref{tab:ablation_map}.

The lowest MSE of MVP is much higher than that of MAP, suggesting that predicting masked moving averages is more feasible than predicting specific masked values in the context of volatile financial data. Moreover, the stock selection performance (IRR and SR) achieved using MAP-based pre-training also surpasses that of MVP-based pre-training under the same experimental settings. Notably, our baseline method~\cite{xia2024ci} is itself designed around masked value prediction and includes specialized structures to improve its performance. Despite this, our SSPT framework outperforms it, further validating the advantages of our MAP design over traditional MVP approaches.

\begin{figure}[b]
\centering
\vspace{-0.4cm}
\begin{center}
   \includegraphics[width=0.98\linewidth]{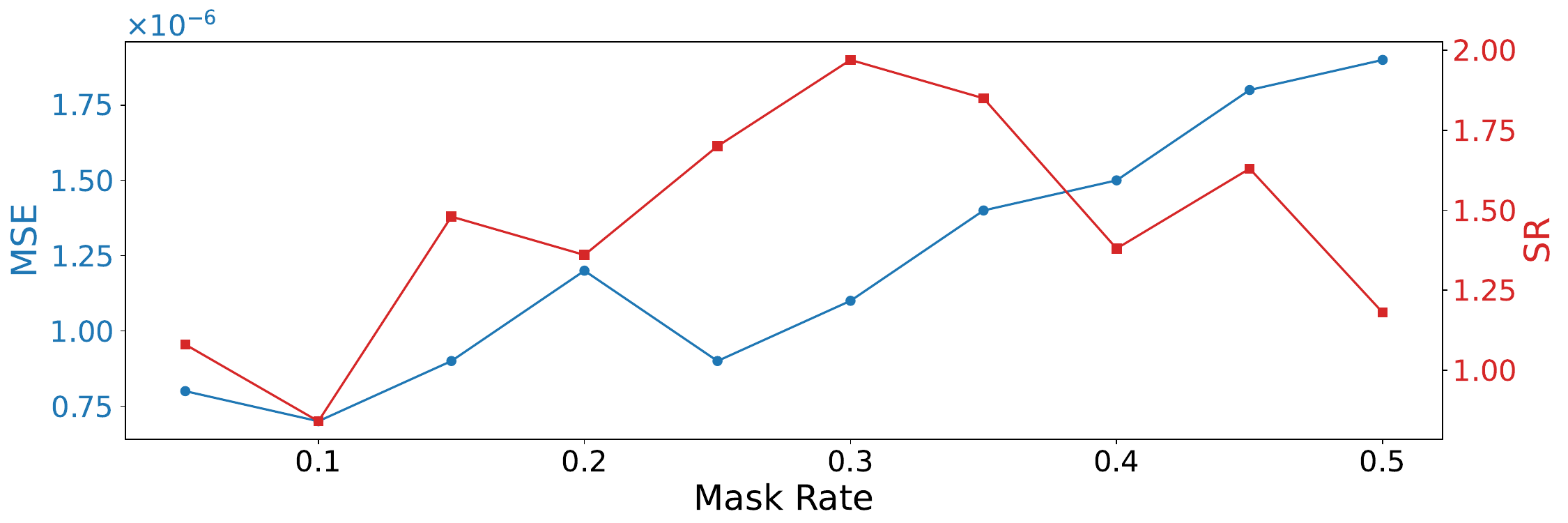}
   \vspace{-0.3cm}
   \captionof{figure}{Pre-training MSE and stock selection SR results using MAP under different mask rate settings. The blue line represents MSE (left y-axis), while the red line indicates SR (right y-axis).}
\label{fig:mask_rate_comparison}
\end{center}
\vspace{-0.0cm}
\end{figure}

Next, we investigate how varying the mask rate affects MAP performance. We vary the mask rate from 0.05 to 0.5 in increments of 0.05 and report the corresponding pre-training MSE and stock selection SR in Figure~\ref{fig:mask_rate_comparison}.

As shown, the MSE increases from $8.0 \times 10^{-7}$ to $1.9 \times 10^{-6}$ as the mask rate increases. In terms of SR, stock selection performance is 1.08 at a mask rate of 0.05, even underperforming the no-pretraining baseline. MAP begins to show benefits at a mask rate of 0.15, with the highest SR observed around a mask rate of 0.3.

\section{IRR Results for Pre-training Analysis}
\label{sec:appendix_IRR_results}
\begin{figure}[t]
\centering
\vspace{-0.0cm}
\begin{center}
   \includegraphics[width=0.98\linewidth]{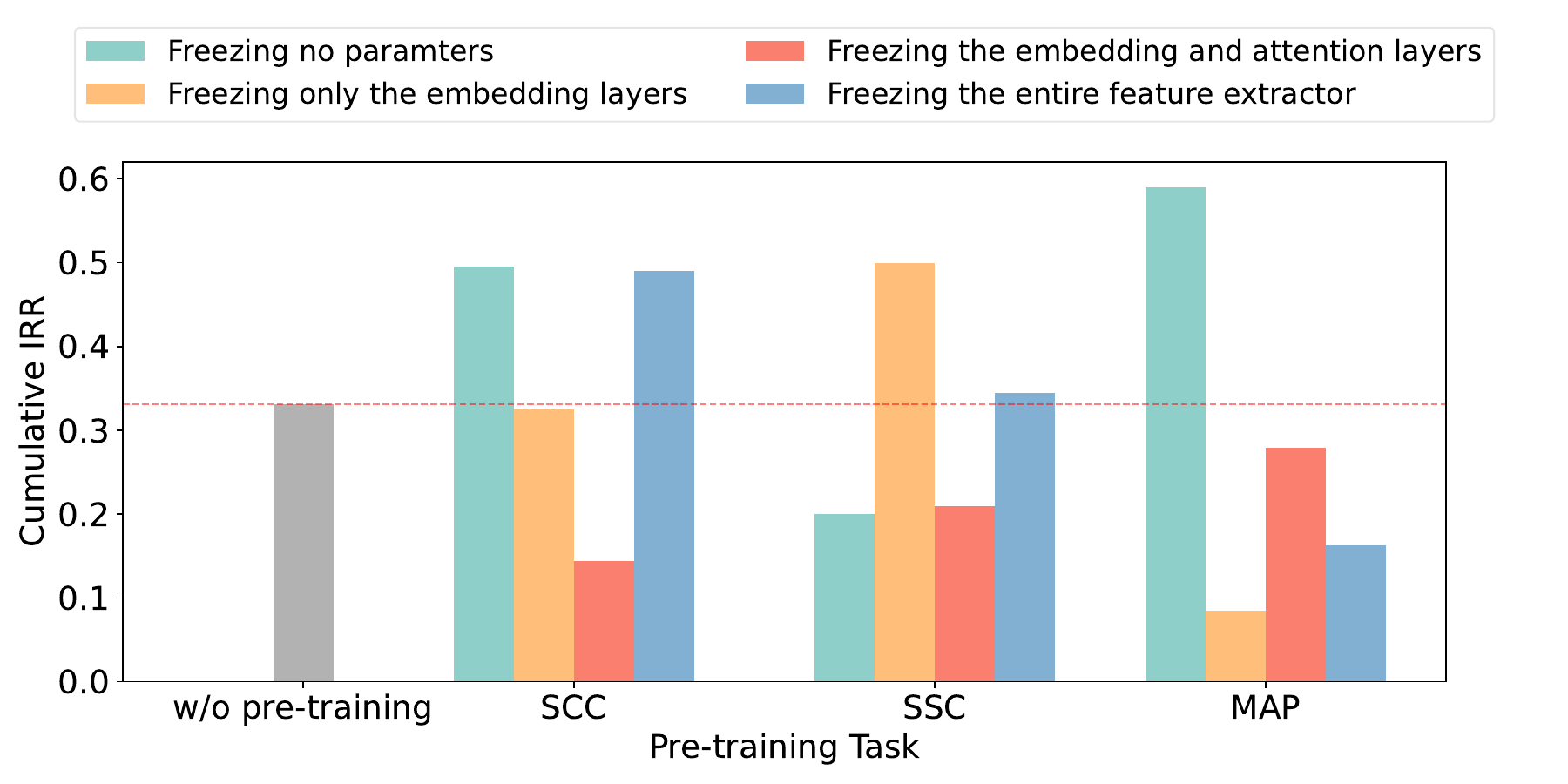}
   \vspace{-0.4cm}
   \captionof{figure}{The cumulative IRR results of the stock selection task with different training strategies. The group `w/o pre-training' presents the result of directly training the model on the stock selection task without any pre-training.}
\label{fig:ft_comparison2}
\end{center}
\vspace{-0.4cm}
\end{figure}

\begin{figure}[t]
\centering
\vspace{-0.0cm}
\begin{center}
   \includegraphics[width=0.98\linewidth]{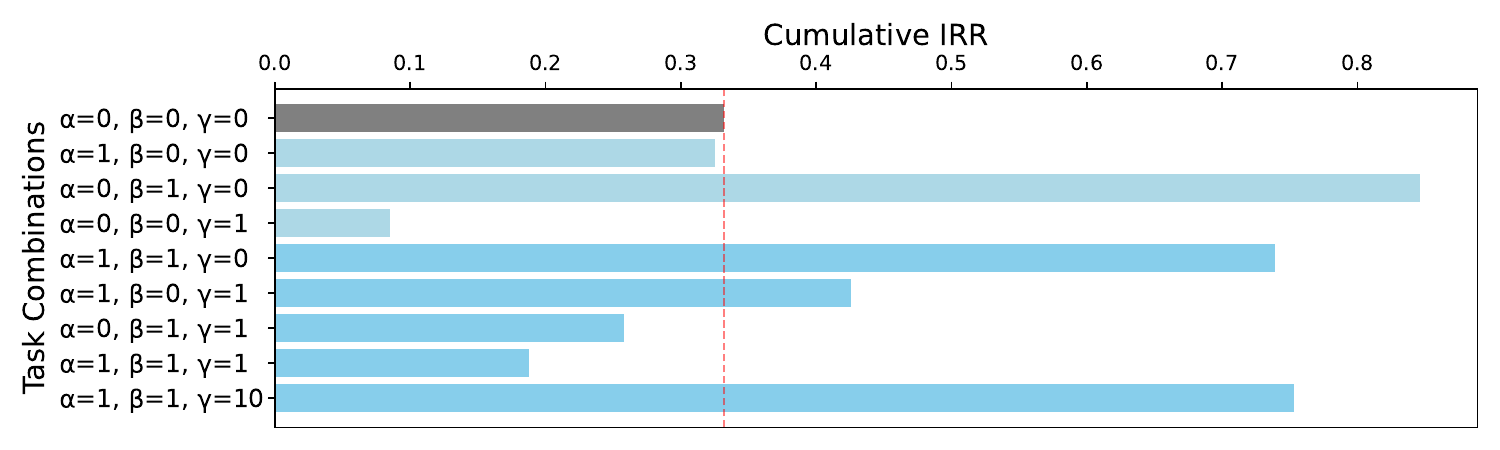}
   \vspace{-0.4cm}
   \captionof{figure}{Cumulative IRR results on the NASDAQ market for the stock selection task using pre-trained models with different combinations of pre-training tasks. The values of $\alpha, \beta, \gamma$ represent the loss coefficients for the SCC, SSC, and MAP tasks in the combined pre-training loss (Equation~\ref{eq:pre-training}). When a coefficient is 0, the corresponding task is excluded from pre-training. The first row shows the baseline result without pre-training. The next three rows display results when pre-training on individual tasks (using the fine-tuning strategy of freezing only the embedding layers). Subsequent rows present results from combinations of pre-training tasks.}
\label{fig:ft_combination2}
\end{center}
\vspace{-0.4cm}
\end{figure}

Both the cumulative investment return ratio (IRR) and Sharpe ratio (SR) are widely used metrics to evaluate stock selection performance. While IRR focuses solely on returns, SR considers both returns and risks. Although these two metrics are not linearly related, their overall trends are generally similar. In the main text, we presented SR results in Figures \ref{fig:ft_comparison} and \ref{fig:ft_combination} to analyze the pre-training tasks. Here, we present the corresponding IRR results in Figures \ref{fig:ft_comparison2} and \ref{fig:ft_combination2}.

If disregarding the specific values, these two figures closely resemble the previous figures for the SR metric. The main difference lies in the fact that the SCC task with frozen embedding layer fine-tuning and the SSC task with frozen embedding and attention layers fine-tuning fail to surpass the performance of direct training without pre-training. However, the most optimized results remain consistent. All conclusions regarding the hyper-parameter and training strategies from the SR metric remain unchanged. This consistency across different metrics indicates the robustness of our methods, demonstrating reliable performance improvements.

\section{More Analysis on Simulated Data}
\label{sec:appendix_simulation}
To further investigate the ability of our classification tasks to capture distinguishing statistical features from time series data, we analyze the sensitivity of the classification accuracy to the degree of difference in the statistical parameters used to generate the simulated series.

\begin{figure}[b]
\centering
\vspace{-0.6cm}
\begin{center}
   \includegraphics[width=0.98\linewidth]{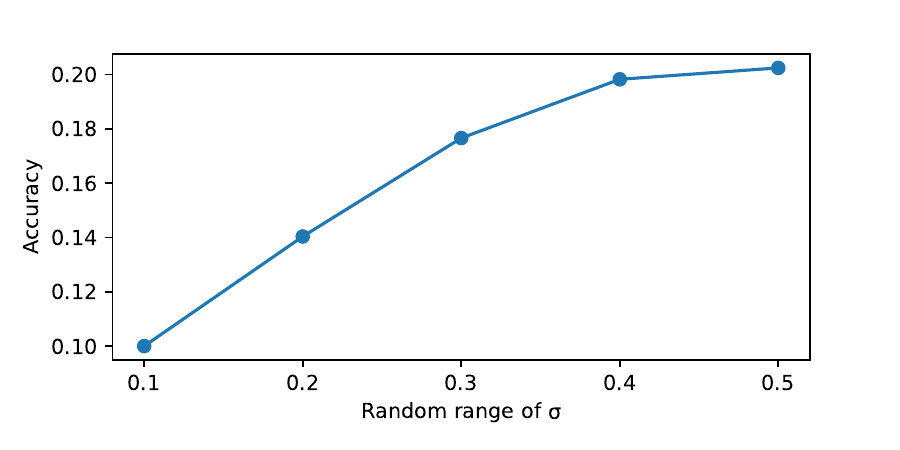}
   \vspace{-0.6cm}
   \captionof{figure}{Classification accuracy on 10 simulated stock price series, where the volatility parameter $\sigma$ is randomly drawn from different ranges. The accuracy is averaged over 5 simulation runs with identical settings for each range of $\sigma$.}
\label{fig:simulation2}
\end{center}
\vspace{-0.0cm}
\end{figure}

Focusing on the classification of 10 simulated series, we control the values of $S(0)$, the time horizon $t$, and the expected return $\mu$ to be the same across all series, while varying the volatility $\sigma$ used to generate each series. Specifically, we change the random range from which the $\sigma$ values are drawn to control the difference in volatilities between the series. For example, $\sigma$ values randomly drawn from $(0.1, 0.2)$, with a range of 0.1, have a smaller difference compared to values drawn from $(0.1, 0.3)$, with a range of 0.2. Figure~\ref{fig:simulation2} presents the classification results for different random ranges of $\sigma$.

We observe a clear upward trend in accuracy as the range of $\sigma$ increases, indicating that the classification task becomes more effective when the statistical features of the time series differ more significantly. This further substantiates our hypothesis that the stock code classification (SCC) and stock sector classification (SSC) tasks can learn to distinguish the unique statistical characteristics inherent to different stocks or sectors, thereby extracting informative features that benefit the subsequent stock selection task.

While practical stock price series exhibit far greater complexity than our simulated data and likely contain additional hidden statistical features, these controlled experiments provide insights into the underlying mechanisms through which our pre-training tasks enhance the stock selection performance. By capturing distinguishing statistical patterns within price series data, the SCC and SSC tasks acquire knowledge that may not directly relate to profitability but can improve the overall understanding and prediction of stock price movements.

\end{document}